\documentclass[conference]{IEEEtran}
\IEEEoverridecommandlockouts
\usepackage{cite}
\usepackage{amsmath,amssymb,amsfonts}
\usepackage{graphicx}
\usepackage{textcomp}
\usepackage{xcolor}
\usepackage{algorithm}
\usepackage{algorithmic}
\usepackage{subfigure}
\usepackage{multicol, multirow}
\usepackage{bbding}
\usepackage{xcolor}         
\usepackage{multirow}
\usepackage{subfigure}
\usepackage{graphicx}
\usepackage{subcaption}
\usepackage{mathrsfs}
\usepackage{tabularray}
\usepackage{amsthm}
\usepackage{booktabs}
\usepackage{amssymb}
\usepackage{hyperref}
\usepackage[misc]{ifsym}

\newtheorem{claim}{Claim}
\def\BibTeX{{\rm B\kern-.05em{\sc i\kern-.025em b}\kern-.08em
    T\kern-.1667em\lower.7ex\hbox{E}\kern-.125emX}}
\begin{document}

\title{A Parameter Update Balancing Algorithm for Multi-task Ranking Models in Recommendation Systems}

\author{
\IEEEauthorblockN{Jun Yuan\textsuperscript{*}}\thanks{* Equal Contribution}
\IEEEauthorblockA{\textit{Huawei Technologies Co., Ltd.} \\
Shenzhen, China \\
yuanjun25@huawei.com}
\and
\IEEEauthorblockN{Guohao Cai\textsuperscript{*}\thanks{\Letter\ Corresponding author}}
\IEEEauthorblockA{\textit{Huawei Noah’s Ark Lab} \\
Shenzhen, China \\
caiguohao1@huawei.com
}
\and
\IEEEauthorblockN{Zhenghua Dong\textsuperscript{\Letter}}
\IEEEauthorblockA{\textit{Huawei Noah’s Ark Lab} \\
Shenzhen, China \\
dongzhenhua@huawei.com} 
}
\maketitle

\begin{abstract}
Multi-task ranking models have become essential for modern real-world recommendation systems. While most recommendation researches focus on designing sophisticated models for specific scenarios, achieving performance improvement for multi-task ranking models across various scenarios still remains a significant challenge. Training all tasks na\"ively can result in inconsistent learning, highlighting the need for the development of multi-task optimization (MTO) methods to tackle this challenge. Conventional methods assume that the optimal joint gradient on shared parameters leads to optimal parameter updates. However, the actual update on model parameters may deviates significantly from gradients when using momentum based optimizers such as Adam, and we design and execute statistical experiments to support the observation. In this paper, we propose a novel Parameter Update Balancing algorithm for multi-task optimization, denoted as \emph{PUB}. In contrast to  traditional MTO method which are based on gradient level tasks fusion or loss level tasks fusion, PUB is the first work to optimize multiple tasks through parameter update balancing. Comprehensive experiments on benchmark multi-task ranking datasets demonstrate that PUB consistently improves several multi-task backbones and achieves state-of-the-art performance. Additionally, experiments on benchmark computer vision datasets show the great potential of PUB in various multi-task learning scenarios. Furthermore, we deployed our method for an industrial evaluation on the real-world commercial platform, \textit{HUAWEI AppGallery}, where PUB significantly enhances the online multi-task ranking model, efficiently managing the primary traffic of a crucial channel. Our code is released at \href{https://github.com/yjdy/Multitask-Optimization-Recommendation-Library}{https://github.com/yjdy/Multitask-Optimization-Recommendation-Library}.



\end{abstract}


\section{Introduction}
Multi-task learning (MTL) is a powerful technique widely applied to various recommendation systems~\cite{DBLP:conf/kdd/MaZYCHC18,DBLP:conf/sigir/MaZHWHZG18,DBLP:conf/recsys/TangLZG20,DBLP:conf/recsys/ZhaoHWCNAKSYC19,DBLP:conf/recsys/LuDS18,clippy23}, including e-commerce, social media, news/video feeds, and online advertising. For instance, in e-commerce, a ranking model can be trained on multiple related tasks, such as predicting user ratings, click-through rate, and purchase likelihood. Similarly, in social media, an MTL model can be trained on tasks such as predicting user engagement with posts, comments, and likes.

\begin{figure}[htp]
    \centering
    \subfigure[Conventional Gradient balancing method (GBM).]{\includegraphics[width=0.38\textwidth]{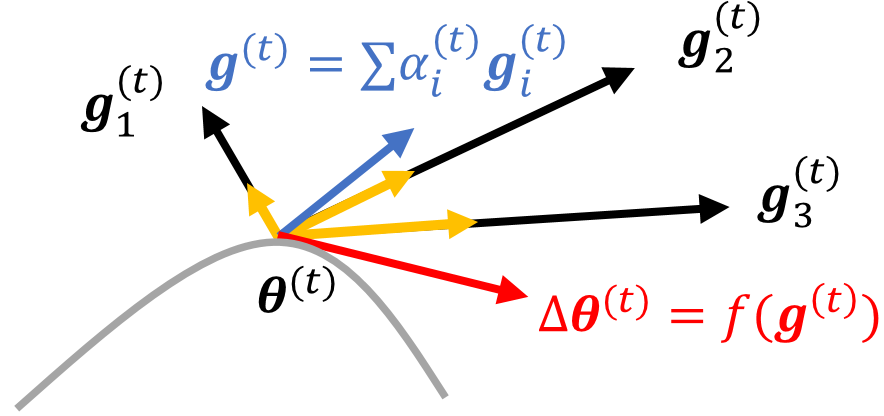}}
    \subfigure[Our parameter update balancing method (PUB).]{\includegraphics[width=0.38\textwidth]{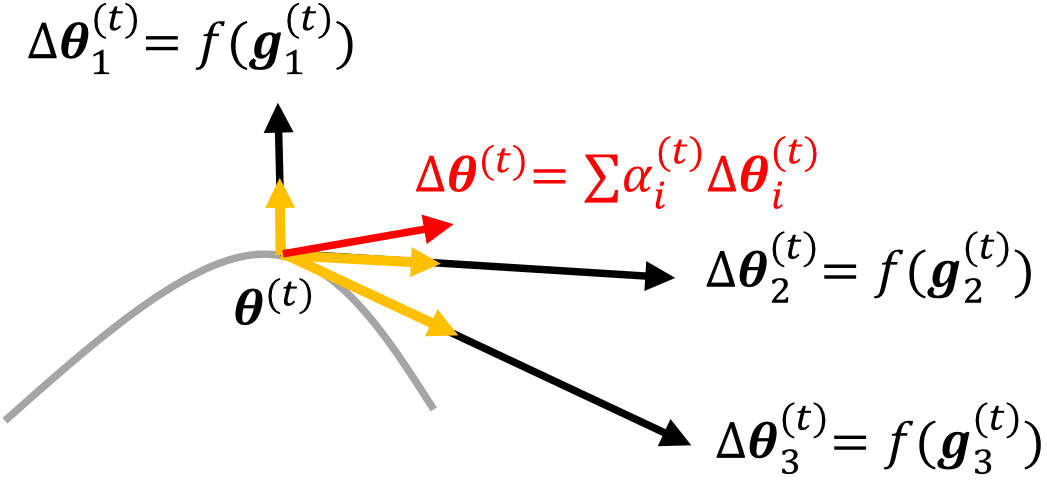}}
    \caption{Visualization of the difference between conventional GBM and PUB. We show the joint gradient obtained by GBMs (blue) and the update of parameters (red). The yellow vectors illustrate the result scaled by the factor $\alpha$, and $f$ denotes the optimizer function.}\label{motivation}
    \vspace{-4mm}
\end{figure}

MTL has demonstrated significant potential for enhancing the accuracy of recommendation systems and user experience. Current multi-task models often share some parameters across all tasks to learn shared representations and then use task-specific parameters to predict multiple labels. However, the intricate and competing correlations among tasks may result in inequitable learning if all tasks are naively trained together. This problem is also known as the \textit{seesaw problem}~\cite{tang2020ple}, where improving performance in one task comes at the expense of other tasks. A major reason behind this issue is conflicting gradients~\cite{liu2021cagrad}, meaning the gradients of different task are not well aligned. As a result, following the average gradient direction can be detrimental to the performance of specific tasks. To address this issue, many multi-task optimization methods have been conducted to alleviate this problem~\cite{mgda,lin2019pareto,Fangrui21pareto,navon2022nashmtl,liu2021imtl,wei23jepoo}. As shown in Figure~\ref{motivation}(a), these methods try to combine per-task gradients (black arrow) into an optimal joint direction (blue arrow). Then, they employ an optimizer to compute parameter updates $\Delta \boldsymbol{\theta}$ (red arrow) to modify model parameters. Although existing MTO methods could alleviate seesaw problem to some extend, they are all based on gradient level tasks fusion or loss level tasks fusion, and weighted loss will converge to gradient balancing with regard to shared parameter updating. It is worth noting that the actual updates of model parameters may deviate significantly from their gradients~\cite{clippy23}, especially when employing momentum based optimizers like Adam and AdaGrad, etc.

\begin{figure*}[htp]
    \centering
    \subfigure{\includegraphics[width=0.2\textwidth,height=0.2\textwidth]{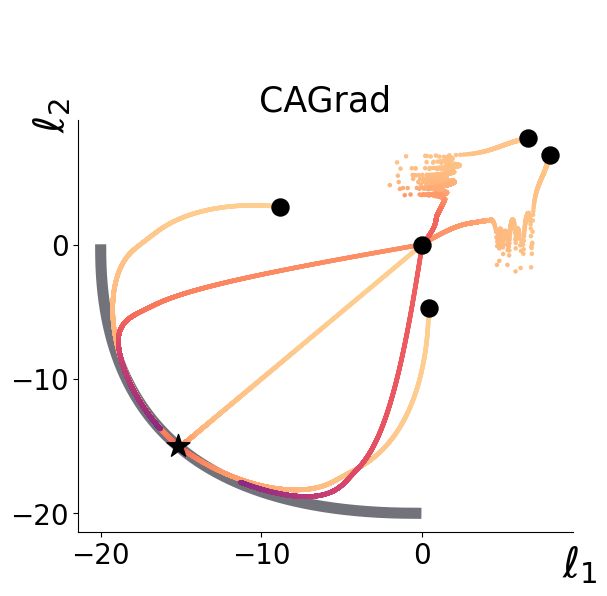}}\hspace{-1mm}\vspace{-1mm}
    \subfigure{\includegraphics[width=0.2\textwidth,height=0.2\textwidth]{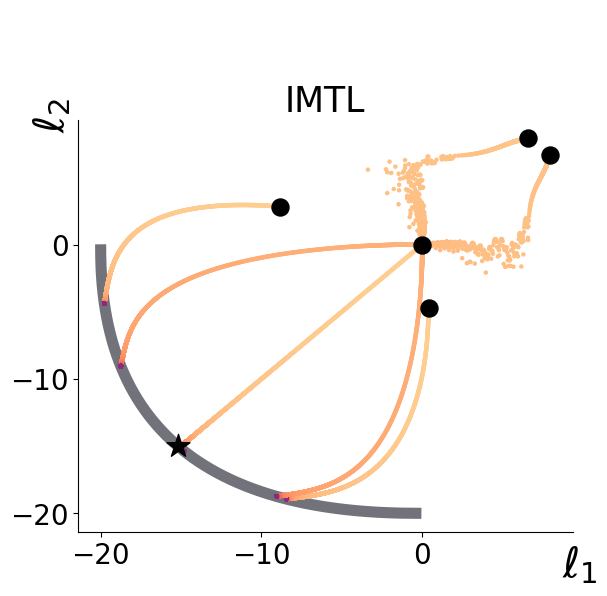}}\hspace{-1mm}\vspace{-1mm}
    \subfigure{\includegraphics[width=0.2\textwidth,height=0.2\textwidth]{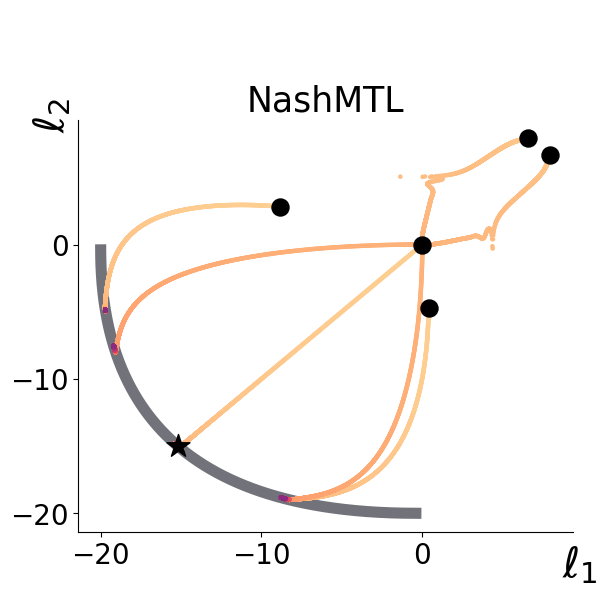}}\hspace{-1mm}\vspace{-1mm}
    \subfigure{\includegraphics[width=0.2\textwidth,height=0.2\textwidth]{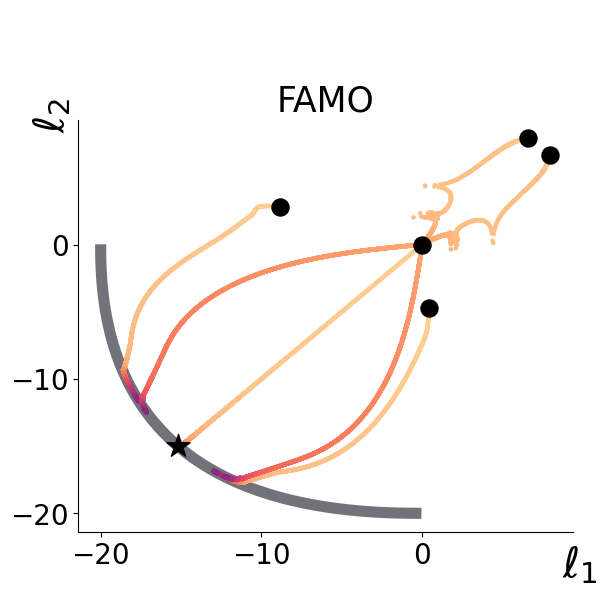}}\hspace{-1mm}\vspace{-1mm}
    \subfigure{\includegraphics[width=0.2\textwidth,height=0.2\textwidth]{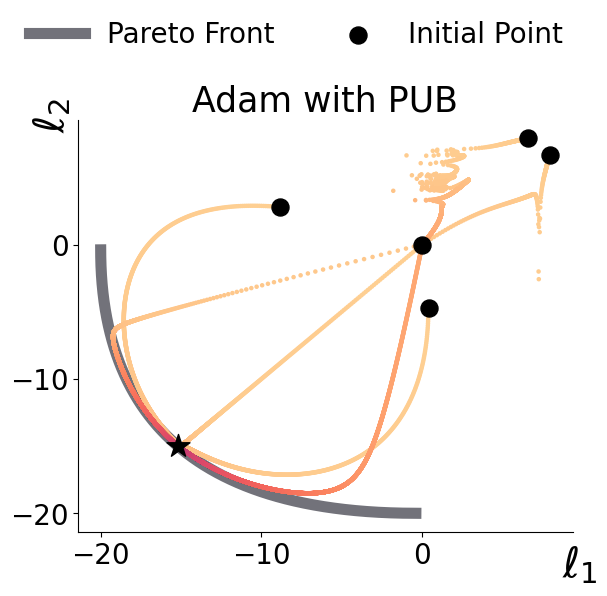}}\hspace{-1mm}\vspace{-1mm}
    \caption{\textit{Toy Experiments}. 
    Optimization trajectories in loss space, and optimizer is Adam in all experiments. Black dots • are 5 different initializations, and their trajectories are colored from orange to purple.  CAGrad, IMTL, NashMTL and FAMO converge to different Pareto-stationary points depending on initial points, while they fail to converge to the optimal solution $\mathcal{L}^{*}$ (star mark). In contrast, PUB  exhibits more robust convergence behavior and can reach the optimal solution for all initializations. Please refer to~\cite{navon2022nashmtl} for details.}\label{fig:toy}
    \vspace{-3mm}
\end{figure*}

In this paper, we propose a Parameter Updating Balancing method, denoted as \emph{PUB}, to overcome the seesaw problem in optimizing multi-task ranking model with shared parameters. As illustrated in Figure~\ref{motivation}(b), unlike traditional methods that balance task gradients, PUB aims to identify the optimal joint parameter update based on task updates. In this way, PUB is able to overcome the limitations of conventional methods mentioned above. As shown in Figure~\ref{fig:toy}, we conducted a toy experiment with Adam~\cite{kingma2014adam} to intuitively understand the differences between PUB and several state-of-the-art (SOTA) methods. Our contributions are summarized as follows:
\begin{itemize}
    \item We identify a fundamental problem with existing gradient balancing methods in benchmark multi-task ranking datasets and further substantiate our observation through the design and execution of statistical experiments in multi-task ranking datasets.
    \item We introduce PUB, a multi-task optimization method that balances parameter updates, as a solution to overcome the limitations of conventional approaches. To the best of our knowledge, PUB is the first work to optimize multiple tasks through update balancing.  
    \item Extensive experiments on four public benchmark ranking datasets demonstrate that PUB improves various multi-task models and achieves state-of-the-art results. Experiments on computer vision dataset show the great potential of PUB in various multi-task learning scenarios. Furthermore, we deployed PUB on a commercial recommendation system, and the online A/B testing results demonstrate the effectiveness of our method.
\end{itemize}

\section{Related Work}

MTL simultaneously learns the tasks by minimizing their empirical losses together. It is common for some specific tasks to be learned well while others are overlooked, \emph{i.e,} the seesaw problem. A main symptom of the seesaw problem is poor overall performance. In this section, we briefly discuss two main categories of MTL methods.

\textbf{Multi-task optimization (MTO)} tries to find task weights that are multiplied by the raw losses for model optimization. Generally, MTO methods can be divided into two categories~\cite{DBLP:journals/corr/ZhangY17aa}, Gradient Balancing and Loss Balancing.

\textit{Gradient Balancing Method (GBM)} attempts to alleviate mutual competition among tasks by combining the per-task gradients of shared parameters into a joint update direction using a particular heuristic. Existing methods typically search for task weights based on a mathematical theory, such as Pareto optimality, which theoretically implies the potential for great generalization of these methodologies~\cite{10.5555/1622248.1622254}. MGDA~\cite{mgda} casts multi-task learning as multi-object optimization and finds the minimum-norm point in the convex hull composed by the gradients of multiple tasks. MGDA-UB~\cite{DBLP:conf/nips/SenerK18} proposes an approximation of the original optimization problem of MGDA to improve efficiency. PCGrad~\cite{Yu20pcgrad} avoids interference between tasks by projecting the gradient of one task onto the normal plane of the other. CAGrad~\cite{liu2021cagrad} optimizes for the average loss while explicitly controlling the minimum decrease rate across tasks. GBMs can evenly learn task-shared parameters while ignoring task-specific ones, making them perform poorly when the loss scale of tasks is huge~\cite{wei23jepoo}. IMTL~\cite{liu2021imtl} and NashMTL~\cite{navon2022nashmtl} are loss scale-free GBM. IMTL-G makes the aggregated gradient have equal projections onto individual tasks, and IMTL-L scales all task-specific loss to be same. NashMTL uses the Nash bargaining solution to make tasks negotiate and reach an agreement on a joint direction of parameter update in the ball of fixed radius centered around zero. The strict restrictions of above two methods make them unable to consistently improve the MTL model.

\textit{Loss Balancing Method (LBM)} uses a delicate heuristic to integrate losses of various tasks based on certain assumptions. Uncertainty weighting~\cite{Kendall2018uncertainty} models the loss weights as data-agnostic task-dependent homoscedastic uncertainty. Then loss weighting is derived from maximum likelihood estimation. GradNorm~\cite{chen2018gradnorm} learns the loss weights to enforce the norm of the scaled gradient for each task to be close. BanditMTL~\cite{mao2021banditmtl} adds the variance of task losses to regularize optimization, but it focuses on the task with a large loss scale. To balance the loss scale, it needs to adopt some tricks. LBMs often have higher efficiency than GBM, and some LBMs, such as GradNorm, can prevent MTL from being biased in favor of tasks with large loss scales.  FAMO~\cite{liu2023famo} is a fast adaptive multi-task optimization method, using a dynamic weighting method that decreases task losses in a balanced way using O(1) space and time. We also compare PUB with several SOTA loss balancing baselines in our experiments.

\textbf{Update manipulation method (UMM)} Besides the MTO methods, there are another type methods called update manipulation method. They limit the updates computed by gradient to prevent issues such as instability, over-fitting in multi-task learning by some heuristics. Popular methods that incorporate UMM include gradient clipping~\cite{10.5555/3042817.3043083,brock2021highperformance}, learning rate warmup~\cite{cohen2022gradient,gilmer2021loss}, and layer normalization~\cite{ba2016layer}. AdaTask~\cite{yang22adatask} utilizes task-specific accumulative gradients when adjusting the learning rate of each parameter, which can alleviate the dominance problem during training. Clippy~\cite{clippy23} tries to mitigate instability by using $\mathcal{L}_{\infty}$ norm on updates instead of gradient, which has been adopted on YouTube multi-task ranking model. Theoretically, PUB is flexible and can integrate UMMs easily. To demonstrate this flexibility, we evaluated AdaTask and Clippy with our proposed method in experiments.

\section{Statistic Analysis}\label{sec:understanding}
In this section, we try to identify the reason why existing MTO methods could not mitigate seesaw problem in some multi-task ranking datasets. We further design statistical experiments to verify our observation of several GBMs in benchmark CTR and CTCVR ranking datasets.

\begin{figure*}[htp]
    \centering
    \subfigure[Share Bottom]{\includegraphics[width=0.32\textwidth]{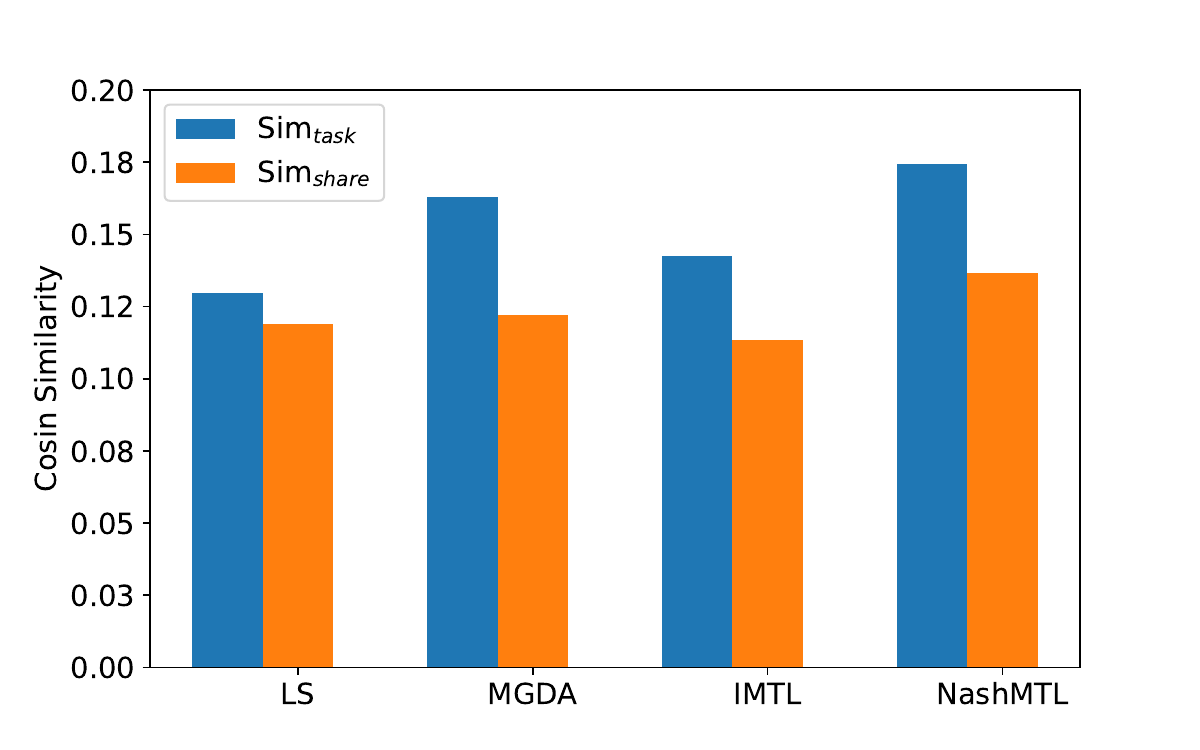}}
    \subfigure[MMOE]{\includegraphics[width=0.32\textwidth]{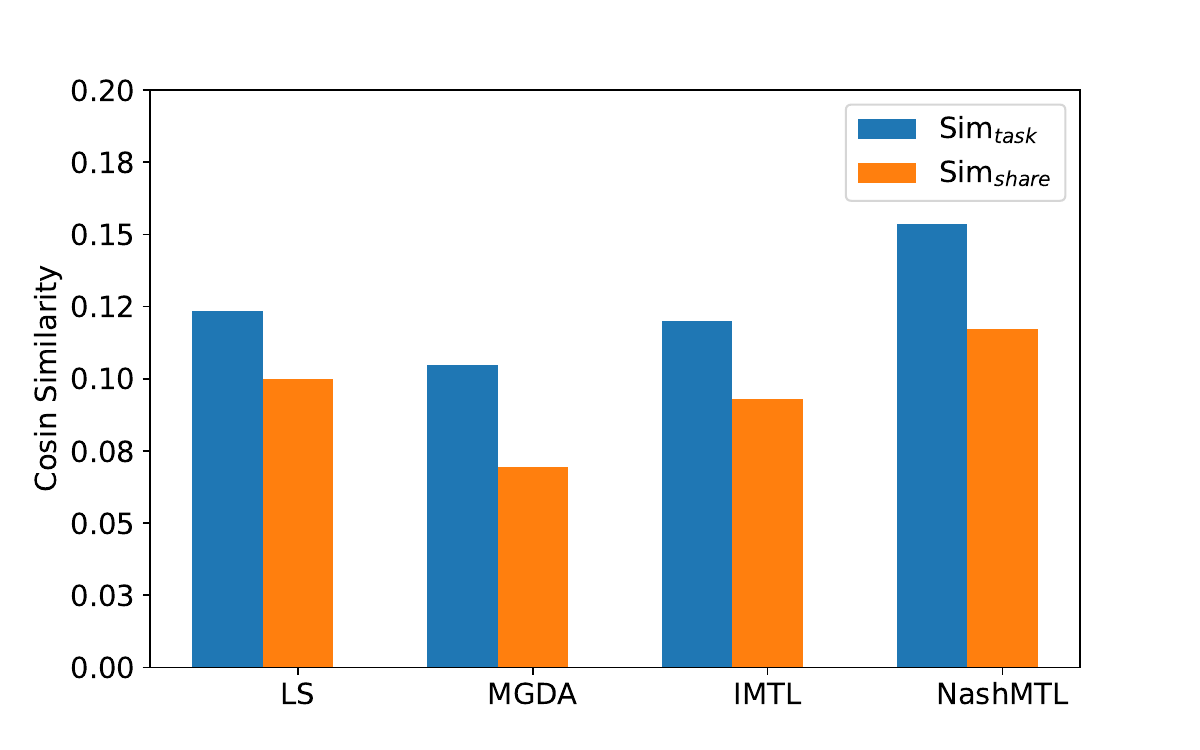}}
    \subfigure[PLE]{\includegraphics[width=0.32\textwidth]{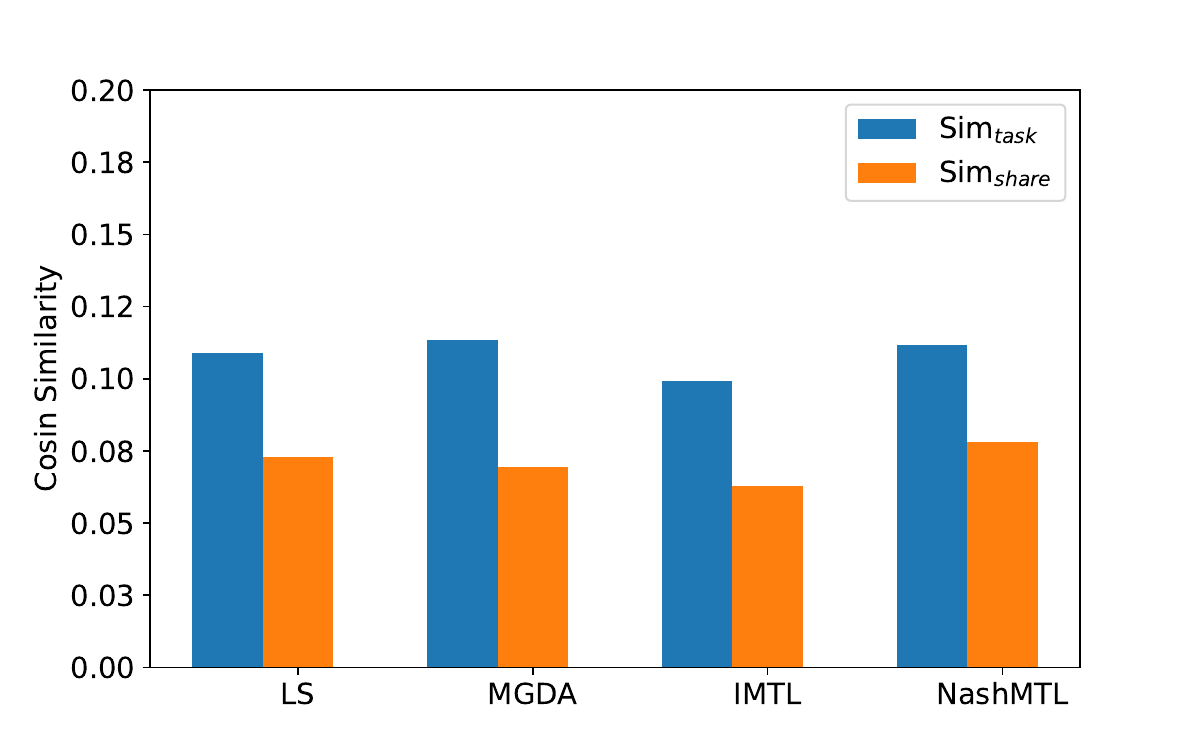}}
    \caption{Cosine similarity between gradient and update of task-specific and shared parameters in AliExpress\_ES dataset. The average cosine similarity of task-specific parameters is significant higher than shared parameters' ($p < 0.01$)}\label{fig:cos_result}
\end{figure*}

\subsection{Root Cause and Evidence}\label{sec:root_cause}
We aim to answer the following research question: (RQ) Does the joint gradient obtained by conventional GBMs leads to the optimal parameter update in multi-task ranking models?

\textbf{Presumption}. We hypothesis that the relationship between gradient and update is relatively optimal when considering only the impact of a single task gradient, such as the relationship within task-specific parameters. Thus, we can compare the relationships between shared parameters and task-specific parameters to determine whether updates on shared parameters are as effective as updates on task-specific parameters.

\textbf{Experiment setting}. We conducted extensive experiments using four different methods and three different MTL models in four country AliExpress datasets (US, NL, ES and FR), which contain two tasks: CTR and CTCVR prediction.
\begin{itemize}
    \item \textbf{MTO methods}: Linear scalarization (LS,$\sum_{i}l_{i}$), MGDA, IMTL-G and NashMTL
    \item \textbf{MTL model}: Shared bottom, MMOE and PLE
    \item \textbf{Optimizer}: Adam
\end{itemize}
We systematically explored all combinations, and tested each setting three times with different random seeds. As a result, 144 experiments are conducted in total. During each experiment, we collected gradient and update of all parameters at every training step from the first step to the end of epoch which get the best overall result in validation dataset. 

A common used metric (called Diff in this paper) is adopted to measure difference between the relationship on task-specific and shared parameters. 
\begin{equation}
    Diff = \frac{2|Sim_{task}-Sim_{share}|}{|Sim_{task}+Sim_{share}|} \label{eq:diff}
\end{equation} where the $Sim_{task}$ and $Sim_{share}$ to represent cosine similarity of gradient and update on task-specific and shared parameters, respectively. $||$ means absolute value. The average cosine similarity results in AliExpress\_ES is shown at Figure~\ref{fig:cos_result}. 

\begin{figure}[htp]
    \centering
    \includegraphics[width=0.35\textwidth]{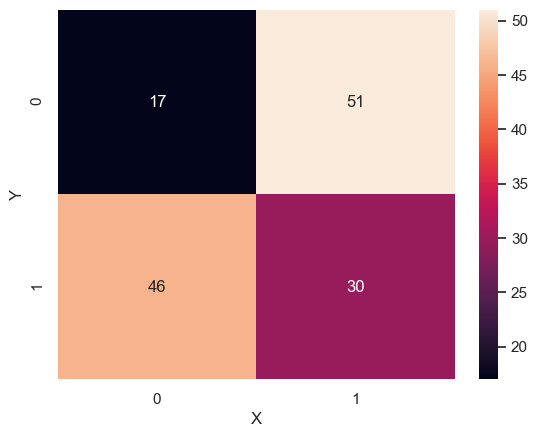}
    \caption{Confusion matrix of Average AUC and Diff metric. We conducted 144 experiments and got 144 pair of $x$ and $y$. A Chi-square test shows a reverse correlation of average AUC and Diff ($p < 0.001$)}\label{confusion_matrix}
\vspace{-2mm}
\end{figure}

\textbf{Analysis.} We try to answer the RQ by revealing the relevance between average AUC and Diff. To begin, we introduce two binary variables $x$ and $y$. The binary indicator $x_{i,j}=0$ means the average AUC of \textit{i}-th experiment is smaller than \textit{j}-th experiment, and vice versa. Similarity, $y_{i,j}=0$ means that the Diff of \textit{i}-th experiment is smaller than the \textit{j}-th experiment, and vice versa. To minimize potential confounding factors, we only compare different methods in the same dataset and using the same model architecture. In short, we obtain a total of 144 $x$, $y$ pairs. The confusion matrix is presented in Figure~\ref{confusion_matrix}.

We employed the Chi-square test of independence to determine if there is a significant association between $x$ and $y$. The test revealed a statistically significant association between the two variables, $\chi^{2}(1) = 16.99, p < 0.001$.  The 67.36\% of the results demonstrate a reverse correlation between average AUC and Diff, \emph{i.e.} lower Diff has high probability of causing higher overall multi-task performance. 

\begin{table}[htp]
\caption{The average cosine similarity of gradient and update on task-specific and shared parameters with different GBMs. We adopt independent samples t-test to compare mean between LS and other methods.}
    \label{tab:cos_sim}
    \centering
    \begin{tabular}{c|c|c|c|c}
    \toprule
         Method&$\overline{Sim}_{task}$&$p$&$\overline{Sim}_{share}$&$p$  \\\hline
         LS&0.130&&0.116&\\
         MGDA&0.138&0.111&0.106&0.011\\ 
         IMTL-G&0.128&0.483&0.102&0.017\\ 
         NashMTL&0.144&0.052&0.1228&0.045\\ 
    \bottomrule
    \end{tabular}
    \vspace{-3mm}
\end{table}

Subsequently, we adopted a t-test to compare $Sim_{task}$ and $Sim_{share}$ between LS and other methods. The average results along with p values are presented in Table~\ref{tab:cos_sim}. The results indicate that there is no significant difference at $Sim_{task}$ between tested GBMs and LS ($p>0.05$). Meanwhile, the average $Sim_{share}$ of MGDA and IMTL-G are significant lower than LS ($p<0.05$), while the average $Sim_{share}$ of NashMTL is significant higher than LS ($p<0.05$). These result demonstrate that the joint gradients obtained by tested GBMs does not lead to updates as effective as updates on task-specific parameters, since tested GBMs cause significant higher Diff than backbones.

\subsection{Statistical Experiment Conclusion}

Our study provides evidence to the following two aspects: 
\begin{enumerate}
    \item It supports our presumption that the relationship between gradient and update is relatively optimal without the impact of multiple task gradients. Our experiments demonstrate a statistically significant reverse association between average AUC and Diff. Specifically, when the relationship between the gradient and update of shared parameters is closer to the relationship on task-specific parameters, there is a high probability of achieving better overall performance.
    \item We challenge the hidden assumption of conventional GBMs in multi-task ranking, namely that (RQ) the optimal joint gradient searched by these methods necessarily leads to optimal update. Because all three methods we tested showed significant differences in the cosine similarity between task-specific and shared parameters.
\end{enumerate}

\section{Preliminary}\label{sec:preliminary}
In this section, some important concepts and notations are introduced for clearly understanding. Below, we summarize some important concepts for Multi-task learning. Additionally, we display some important notations in Table~\ref{tab:notations}.

\begin{table}[htp]
\caption{Summary of notations}
\label{tab:notations}
\centering
\scalebox{0.95}{
\begin{tabular}{c|c}
\toprule
\textbf{Notation}&\textbf{Description}\\\hline
$T$, $t$ &Total and current training step\\
$n$&Number of the including tasks\\
$\boldsymbol{\alpha}$&Task weights: combination weight vector for task gradients \\
$f(\cdot)$ & Optimizer function \\
$\boldsymbol{g}_{i}$&Task gradient: gradient generated only by loss of task $i$ \\
$\Delta \boldsymbol{\theta}_{i}$& Parameter updates computed only by $\boldsymbol{g}_{i}$, \emph{i.e,} $\Delta \boldsymbol{\theta}_{i} = f(\boldsymbol{g}_{i})$\\
$\mathcal{L}_{i}$ & Loss of \textit{i}th task, or simply task loss i\\
$s(\cdot)$ & Parameter updating manipulation method \\
\bottomrule
\end{tabular}}
\vspace{-3mm}
\end{table}



\textbf{Pareto Optimal and Pareto front}. Pareto optimality is a situation where no action or allocation is available that makes one individual better off without making another worse off\footnote{https://en.wikipedia.org/wiki/Pareto\_efficiency}. A solution $x$ dominates $x'$ if it is better on one or more objectives and not worse on any other objectives. A solution that is not dominated by any other is called Pareto optimal, and the set of all such solutions is called the Pareto front. How to identify Pareto front is called Pareto optimal problem.



\section{Method}\label{sec:method}
Based on Section~\ref{sec:understanding}, we know that without gradients of other tasks, update obtained by task gradient ($i.e. \Delta \boldsymbol{\theta}_{i} = f(\boldsymbol{g}_{i})$) is sufficient for updating parameters. This section details PUB, an efficient method for achieving optimal updates. We first provide a general overview of the PUB algorithm, followed by a formalization of the update balancing problem as a convex optimization issue. Finally, we outline a method for efficiently approximating this convex optimization problem. 

\subsection{PUB Algorithm}

\textbf{Utility Definition.} We define the utility function for each task as $u_{i}(\theta) = (\Delta \theta_{i})^\top \Delta\theta$. $\forall i, u_{i} \in \mathbb{R}$. Intuitively speaking, higher utility $u_{i}(\theta)$ of task $i$ means better optimization for the task at current training step. Thus, each task desires to maximize its utility at any training step. That is to say, avoiding seesaw problem means improving utility of one task without decreasing utilities of other tasks. In this way, parameter update balancing problem can be seen as a Pareto optimal problem (as mentioned in Section~\ref{sec:preliminary}), and the optimal solution must on Pareto front.

\textbf{Problem Formalization.} The parameter update balancing problem can be formalize as finding weight vector $\boldsymbol{\alpha}$ that satisfies $\boldsymbol{D}^\top \boldsymbol{D}\boldsymbol{\alpha} = 1/{\boldsymbol{\alpha}}$, where $\boldsymbol{D}$ is the $d \times n$ matrix whose columns are the task parameter updates $\Delta \theta_{i}$, where $d$ is shared parameters number and $n$ is task number. The related proof please refer to subsection~\ref{subsec:proof}.

\textbf{Algorithm} In Alg~\ref{alg_adam}, we shown the Adam~\cite{kingma2014adam}, AdaGrad~\cite{jonh11adagrad} and RMSProp with PUB, respectively. The pseudo code only demonstrates procedures for shared parameters, as the procedures for task-specific parameters are identical for all optimizers. The detailed description is as follow:

Firstly, PUB calculates task gradients by computing the backward partial derivatives of the task losses with respect to the model parameters separately: $\boldsymbol{g}_{i}^{(t)} = \nabla_{\boldsymbol{\theta}^{(t)}} \mathcal{L}_{i}^{(t)}$. Secondly, PUB uses the adopted UMM to compute task updates and obtain the matrix $\boldsymbol{D}$. It should be noted that PUB is a flexible approach, as people can choose to use different UMMs or not, depending on the specific situation. Thirdly, the optimal task weights $\boldsymbol{\alpha}$ are obtained using a sequential optimization approach described in subsection~\ref{subsec:method}. Next, PUB combines the task updates with $\boldsymbol{\alpha}$ to form the optimal joint update direction and updates the shared parameters. Lastly, PUB updates the moment vectors after updating parameters, as this can make training more stable, according to~\cite{clippy23}. 

When updating the momentum, we simply use the same weights as aggregating updates. On the one hand, this trick can reduce complexity and save training time; on the other hand, it does not appear to have a  negative impact in our experiments. We leave a theoretical analysis of this trick in the future work. In our practice, PUB updates moment vectors after updating parameters, but people can seem it as a hyperparameter and tune it according to actual situation.

\begin{algorithm}[tp]
	\caption{AdaGrad, RMSProp and Adam with PUB} \label{alg_adam}
\vspace{-1.5mm}
	\begin{algorithmic}[1]
		\REQUIRE  learning rate $\eta$, $\{\beta\}$ and \emph{eps}. Initial parameters: $\boldsymbol{\alpha}^{0} = (1, 1,..., 1)$, randomly initialize shared parameters $\theta^{(0)}$, update manipulation method $s(\cdot)$
		
		\FOR {$t \leftarrow 1$ to $T$}
            
            \FOR {$i \leftarrow 1$ to $n$}
                \STATE Compute the gradient of every tasks on shared parameters $\boldsymbol{g}_{i}^{(t)} = \nabla_{\boldsymbol{\theta}^{(t)}} \mathcal{L}_{i}^{(t)}$
                \STATE Compute the updates of every tasks on shared parameters 
                \IF{Adam}
                \STATE $\Delta \boldsymbol{\theta}^{(t)}_{i} = \frac{\beta_{1}\boldsymbol{m}^{(t)}+(1-\beta_{1})\boldsymbol{g}_{i}^{(t)}}{\sqrt{\beta_{2}\boldsymbol{v}^{(t)}+(1-\beta_{2})(\boldsymbol{g}_{i}^{(t)})^{2}}+eps}$
                \ENDIF
                \IF{AdaGrad or RMSProp}
                \STATE $\Delta \boldsymbol{\theta}^{(t)}_{i} = \frac{\boldsymbol{g}^{(t)}_{i}}{\sqrt{\beta \boldsymbol{G}^{(t)}+(1-\beta)(\boldsymbol{g}^{(t)}_{i})^{2}}+eps}$
                \ENDIF
            \ENDFOR
            \STATE Use UMM $s(\cdot)$ and computing updates matrix,
            $\boldsymbol{D}^{(t)} = [s(\Delta \boldsymbol{\theta}^{(t)}_1),...,s(\Delta \boldsymbol{\theta}^{(t)}_{n})]$. 
            \STATE And solve for $\alpha$: 
            $(\boldsymbol{D}^{(t)})^\top \boldsymbol{D}^{(t)}\boldsymbol{\alpha}^{(t)} = 1/{\boldsymbol{\alpha}^{(t)}}$
            \STATE Update shared parameter $\theta$:
            $\theta^{(t+1)} = \theta^{(t)} - \eta \boldsymbol{D}^{(t)} \boldsymbol{\alpha}^{(t)}$
            \STATE Update moment vector
            \IF{Adam}
            \STATE $\boldsymbol{m}^{(t+1)} = \beta_{1}\boldsymbol{m}^{(t)}+(1-\beta_{1})\sum_{i}^{n}\alpha^{(t)}_{i}\boldsymbol{g}_{i}^{(t)}$
            \STATE $\boldsymbol{v}^{(t+1)} = \beta_{2}\boldsymbol{v}^{(t)}+(1-\beta_{2})(\sum_{i}^{n}\alpha^{(t)}_{i}\boldsymbol{g}_{i}^{(t)})^{2}$
            \ENDIF
            \IF{AdaGrad or RMSProp}
            \STATE $\boldsymbol{G}^{(t+1)} = \beta \boldsymbol{G}^{(t)} +(1-\beta)(\sum_{i}^{n} \alpha_{i}^{(t)} \boldsymbol{g}^{(t)}_{i})^{2}$
            \ENDIF
        \ENDFOR
		\RETURN $\theta$ with best validation performance
	\end{algorithmic}
  \vspace{-1.5mm}
\end{algorithm}

Theoretically, PUB can be integrated with different optimizers and UMMs easily. Because above analysis does not make any assumption or constrain on the method to get $\Delta \theta_{i}$.  PUB only has two differences when integrating different optimizers, computing task updates $\Delta \theta_{i}$ and updating moment vectors. To demonstrate the flexibility of PUB in a low cost way, we have conducted experiments on PUB with Adam and some UMMs and left the extension research for future work.

\subsection{Formalizing PUB as Convex Problem}\label{subsec:proof}
Next, we formalize parameter update balancing as a convex problem. We provide the proof for our claim that the solution of update combination step is same as $D^\top D\alpha = 1/{\alpha}$. Given shared parameters $\theta$ of a MTL model, we aim to find an update vector $\Delta \theta^{(t)}$ at training step $t$ within the ball of radius $\epsilon \in B_{\epsilon}$ (omitting superscript t in the following). We presume that if $\theta$ is not Pareto stationary then the task updates are linearly independent~\cite{navon2022nashmtl}. Therefore, if $\theta$ is not on the Pareto front, the unique solution has the following form:

\begin{claim} Let $\boldsymbol{D}$ be the $d \times n$ matrix whose columns are the task parameter updates $\Delta \theta_{i}$, where $d$ is shared parameters number and $n$ is task number. The solution to $\arg \max_{\Delta \theta \in B_{\epsilon}} \sum_{i} \log(\Delta \theta^\top \Delta\theta_{i})$ is the solution to $\boldsymbol{D}^\top \boldsymbol{D}\boldsymbol{\alpha} = 1/{\boldsymbol{\alpha}}$ where $\boldsymbol{\alpha} \in \mathbb{R}^{T}_{+}$ and $1/{\boldsymbol{\alpha}}$ is the element-wise reciprocal.
\end{claim}

Our proof assume that:
1) without gradient of other tasks, update obtained by task gradient ($i.e. \Delta \boldsymbol{\theta}_{i} = f(\boldsymbol{g}_{i})$ where $f$ is the optimizer) is optimal for the task. 2) if $\theta$ is not Pareto stationary then the gradients are linearly independent \cite{navon2022nashmtl}. It should be noted that when computing $\Delta \theta_{i}$ the moments of optimizer are constants, making $\Delta \theta_{i}$ independent as well. 

\begin{proof}
    The derivative of this objective is $\sum_{i=1}^{T} \frac{1}{\Delta \theta^\top \Delta\theta_{i}}\Delta\theta_{i}$. For all vector $\Delta\theta$ such that $\forall i: \Delta\theta^\top \Delta\theta_{i} > 0$ the utilities are monotonically increasing with the norm of $\Delta\theta$. Thus, the optimal solution has to be on the boundary of $B_{\epsilon}$. From this we see that the update at the optimal point $\sum_{i=1}^{T} \frac{1}{\Delta \theta^\top \Delta\theta_{i}}\Delta\theta_{i}$ must be in the radial direction, \emph{i.e.} $\sum_{i=1}^{T} \frac{1}{\Delta \theta^\top \Delta\theta_{i}}\Delta\theta_{i} = \lambda \Delta\theta$. Since we have $\Delta \theta = \sum_{i}\alpha_{i}\Delta\theta_{i}$, and the updates of different tasks are independent. Thus, we must have $\sum_{i=1}^{T} \frac{1}{\Delta \theta^\top \Delta\theta_{i}}\Delta\theta_{i} = \sum_{i}\lambda\alpha_{i}\Delta\theta_{i}$. As the inner product must be positive for a descent direction we can conclude $\lambda > 0$; we set $\lambda = 1$ to ascertain the direction of $\Delta \theta$ (the norm might be larger than $\epsilon$). Now finding the bargaining solution is reduced to finding $\alpha \in \mathbb{R}^{T}$ with $\alpha_{i} > 0$ such that $\forall i: \Delta\theta^\top \Delta\theta_{i} = \sum_{j}\alpha_{j}\theta_{j}^\top \theta_{i} = 1/{\alpha_{i}}$. This equivalent to requiring that $D^\top D\alpha=1/\alpha$ where $1/\alpha$ is the element-wise reciprocal.
\end{proof}

\subsection{Efficient approximate method}\label{subsec:method}
Thirdly, we design a method through a sequence of convex optimization problems to efficiently approximate the optimal solution for $D^\top D\alpha = 1/\alpha$. 

To begin, we define $\beta_{i}(\alpha) = \Delta\theta_{i}D\alpha$. Ideally, if we get optimal solution $\alpha$, we can obtain $\beta_{i}$ for all $i$ by  $\beta_{i} = 1/\alpha_{i}$. Equivalently, we can express $\log(\alpha_{i}) + \log(\beta_{i}) = 0$. Therefore, we introduce the following function, and our objective is to find a non-negative $\alpha$ such that $\varphi_{i}(\alpha)$
\begin{equation}
    \varphi_{i}(\alpha) = \log(\alpha_{i}) + \log(\beta_{i})
\end{equation}

To get optimal solution $\alpha$, we consider the following optimization problem:
\begin{equation}
\begin{aligned}
    &\min_{\alpha} \sum_{i} \varphi_{i}(\alpha) \\
    s.t. \forall i, &-\varphi_{i}(\alpha) \leq 0 \\
    &\alpha_{i} > 0
\end{aligned}
\end{equation}

Next, we employ an iterative first-order optimization algorithm to tackle the aforementioned optimization problem more efficiently. Specifically, we replace the concave term $\varphi_{i}(\alpha)$ with its first-order approximation $$\tilde{\varphi}_{i}(\alpha^{(\tau)}) = \varphi_{i}(\alpha^{(\tau)}) + \nabla \varphi_{i}(\alpha^{(\tau)})^\top (\alpha - \alpha^{(\tau)})$$ where $\alpha^{(\tau)}$ is the solution at iteration $\tau$. It is worth noting that we only modify $\varphi$ to $\tilde{\varphi}$ in the objective function, while keeping the constraint on $\varphi$, resulting in an optimized problem can be rewritten as follows:
\begin{equation}
\begin{aligned}
    &\min_{\alpha} \sum_{i} \tilde{\varphi}_{i}(\alpha) \\
    s.t. \forall i, &-\varphi_{i}(\alpha) \leq 0 \\
    &\alpha_{i} > 0
\end{aligned}
\end{equation}

This sequential optimization approach is a variant of the concave-convex procedure (CCP)~\cite{10.1162/08997660360581958,Lipp2016VariationsAE}, so we can leverage existing tools, such as CVXPY~\cite{diamond2016cvxpy}, to get the critical points of $\alpha^{(\tau)}$ for every step $\tau$. Furthermore, since we do not modify the constraint, $\alpha^{(\tau)}$ always satisfies the constraint of original problem for any given step $\tau$. 

\textbf{Complexity Analysis}. According to the page 48 of~\cite{bubeck2015convex}, we would like to report the complexity of PUB as follows: If we compute the value of parameter $\alpha$ in every training step, there will be $O((n+\tau)d+\tau^2)$ additional computation, where $n$ is the number of tasks, $\tau$ is the iteration steps with the maximum 200 in our experiments, $d$ is the number of shared parameters. In practice, we compute the value of parameter $\alpha$ in every $s \in \{10, 100\}$ training steps to accelerate training, and the total training time cost is lower than the compared methods.

\section{Experiments}\label{sec:exp}

\subsection{Experimental Setup}
We conducted experiments in two publicly benchmark datasets and deployed our PUB algorithm in a commercial platform. The public datasets contains both CTR\&CTCVR prediction and scene understanding.
\subsubsection{Public Datasets}\label{datasets}
For \textit{CTR\&CTCVR prediction}, we experiment in the AliExpress Dataset\footnote{https://tianchi.aliyun.com/dataset/74690}. This dataset collects user logs from the real-world traffic in the AliExpress e-commercial platform. We use data from four countries: the Netherlands, Spain, France and the USA. It contains two binary classification tasks: Click Through Rate (CTR) prediction and post-click conversion rate (CTCVR) prediction. \textit{For scene understanding}, we follow the protocol of~\cite{liu2019mtan,navon2022nashmtl} in the NYUv2 dataset~\cite{nyuv2}. It is an indoor scene dataset that consists of 1449 RGBD images, each with dense per-pixel labeling and 13 classes. It contains three tasks,  semantic segmentation, depth estimation, and surface normal prediction. We use the dataset as a three tasks learning benchmark dataset.

\subsubsection{MTL architecture}\label{backbones}
(1) MMOE~\cite{ma2018mmoe} extends MOE to utilize different gates for each task to obtain different fusing weights in MTL. (2) PLE~\cite{tang2020ple} develops both shared experts and task-specific experts, together with a progressive routing mechanism to further improve learning efficiency. In our experiments, all experts are single layer perceptron with same shape. MMOE backbone has two moe layers which both contain 8 experts. PLE has two moe layers too, and each moe layer contains 8 shared experts and 8 task-specific expert for every tasks. (3) Task Attention Network (MTAN)~\cite{liu2019mtan} is the architecture in scene understanding experiment. It adds an attention mechanism on top of the SegNet architecture~\cite{Badrinarayanan2017SegNet}.

\subsubsection{Implementation Details and Evaluation Metrics}

For ranking datasets AliExpress, we trained two popular multi-task recommendation backbones, \emph{i.e.} MMOE~\cite{ma2018mmoe} and PLE~\cite{tang2020ple} in all datasets using two independent tower with the multi-expert layer. The MMOE model contain eight shared single layer fully-connected network as expert networks, while the PLE model contains four shared expert networks and four task-specific expert networks for each task. We re-implemented Uncertainty, MGDA, NashMTL, BanditMTL, IMTL, CAGrad, PCGrad and FAMO, based on the open-source code\footnote{https://github.com/Cranial-XIX/FAMO} of \cite{liu2023famo}. The learning rate was set to 1e-3 and weight decay to 1e-6, with a batch size of 4096. We adopt AUC~\cite{guo17deepfm} as the evaluation metrics, which is commonly used in CTR and CTCVR prediction. \textbf{AUC} measures the probability of a positive sample being ranked higher than a randomly chosen negative one. We also report average AUC of CTR and CTCVR as the overall performance. 

\begin{table*}[htp]
\caption{Two Tasks Experimental Result in multi-task ranking dataset Aliexpress. AUC Values are averages over 5 random seeds and only keep 4 decimal. Average is the arithmetic mean of CTR and CTCVR results, and we use the metric to measure overall performance. The best results in each column are bold. The $^\ddag$ indicates the result is a significantly improvement than backbone with $p<0.05$}
\label{tab:two_tasks}
\centering
\scalebox{0.92}{
\begin{tabular}{l|ccc|ccc|ccc|ccc}
\toprule
\multirow{2}{*}{\textbf{Method}}& \multicolumn{3}{c|}{US} &\multicolumn{3}{c|}{ES} & \multicolumn{3}{c|}{FR} & \multicolumn{3}{c}{NL} \\
& CTR & CTCVR & Average & CTR & CTCVR & Average & CTR & CTCVR & Average & CTR & CTCVR & Average\\\hline
Single Task Learning &0.7058 &0.8637 &0.7848 &0.7252 & 0.8832&0.8042 & 0.7174 &0.8702&0.7938 & 0.7203 & 0.8556 &0.7880\\\hline
\textbf{MTL Model}& & & & & & & & &&&& \\
MMOE&0.7092&0.8717&0.7905&0.7290&0.8959&0.8124&0.7274&0.8789&0.8032&0.7247&0.8645&0.7946\\
PLE &0.7071&0.8732&0.7901&0.7306&0.8930&0.8118&0.7297&0.8780&0.8038&0.7233&0.8639&0.7936\\\hline
\textbf{GBM}& & & & & & & &&&&&\\
MMOE+MGDA &0.7060 &0.8703 &0.7882 &0.7294&0.8901 &0.8098 &0.7279 &0.8766&0.8022&0.7231&0.8608&0.7920\\
PLE+MGDA &0.7066&0.8674&0.7870&0.7278 &0.8928 &0.8103 &0.7263 &0.8806$^\ddag$&0.8034&0.7249$^\ddag$&0.8583&0.7916\\
MMOE+NashMTL &0.7088&0.8723&0.7905&0.7285&0.8957&0.8121&0.7241&0.8777&0.8009&0.7255&0.8616&0.7935\\
PLE+NashMTL &0.7067 &0.8719&0.7894&0.7292&0.8929&0.8111&0.7228&0.8794$^\ddag$&0.8011&0.7262&0.8642&0.7952$^\ddag$\\ 
MMOE+IMTL-G &0.7026&0.8643&0.7835&0.7275&0.8922&0.8098&0.7226&0.8742&0.7984&0.7237&0.8609&0.7923\\
PLE+IMTL-G &0.7087$^\ddag$&0.8707&0.7897&0.7304&0.8897&0.8101&0.7255&0.8796&0.8026&0.7256$^\ddag$&0.8645&0.7950$^\ddag$\\
MMOE+PCGrad &0.7092&0.8684&0.7888&0.7301$^\ddag$&0.8915&0.8108&0.7278&0.8788&0.8033&0.7244&0.8620&0.7932\\
PLE+PCGrad &0.7101$^\ddag$&0.8634&0.7868&0.7295&0.8930&0.8113&0.7264&0.8792&0.8028&0.7266&0.8623&0.7945\\
MMOE+CAGrad&0.7080&0.8720&0.7899&0.7296&0.8934&0.8115&0.7257&0.8784&0.8020&0.7232&0.8575&0.7903\\
PLE+CAGrad &0.7083$^\ddag$&0.8725&0.7903&0.7304&0.8950$^\ddag$&0.8127&0.7281&0.8782&0.8032&0.7236&0.8604&0.7920\\\hline
\textbf{LBM}& & & & &  &&&&& & &\\
MMOE+Uncertainty &0.7050&0.8670&0.7860&0.7286&0.8932&0.8109&0.7248&0.8779&0.8013&0.7245& 0.8614&0.7930\\
PLE+Uncertainty &0.7061&0.8670&0.7866&0.7291&0.8932&0.8112&0.7258&0.8784&0.8021&0.7241& 0.8622&0.7932\\
MMOE+BanditMTL &0.7073& 0.8698&0.7885&0.7292&0.8949&0.8120&0.7284$^\ddag$&0.8781&0.8033&0.7252& 0.8610&0.7931\\
PLE+BanditMTL &0.7068& 0.8678& 0.7873&0.7292& 0.8936& 0.8114&0.7264&0.8781&0.8023&0.7241& 0.8605&0.7923\\
MMOE+FAMO &0.7085&0.8626&0.7855&0.7295&0.8941&0.8118&0.7239&0.8709&0.7934&0.7266&0.8596&0.7931\\
PLE+FAMO &0.7082&0.8581&0.7831&0.7307&0.8870&0.8089&0.7266&0.8710&0.7988&0.7245&0.8600&0.7923\\
\hline
\textbf{UMM}& & & & & & &&&&& &\\
MMOE+ClippyAdam&0.7100&0.8694&0.7897&0.7303$^\ddag$&0.8931&0.8117&0.7285$^\ddag$&0.8735&0.8010&\textbf{0.7267}$^\ddag$&0.8627&0.7947\\ 
MMOE+AdaTask&0.7095 &0.8688 &0.7891 &0.7306$^\ddag$ &0.8917 &0.8111 &0.7249&0.8776&0.8012&0.7241&0.8627 &0.7934\\ 
PLE+ClippyAdam&0.7083$^\ddag$ & 0.8654&0.7868 &0.7311 &0.8918&0.8114&0.7253&0.8779&0.8016&0.7228&0.8601 &0.7914\\ 
PLE+AdaTask&0.7083$^\ddag$ &0.8650 &0.7866 &0.7315 &0.8929 &0.8122 &0.7249&0.8790$^\ddag$&0.8020&0.7237&0.8613&0.7925\\\hline
\textbf{Ours}& & & & & & &&&&& &\\
MMOE+PUB & 0.7121$^\ddag$&	\textbf{0.8736}$^\ddag$	&\textbf{0.7929}$^\ddag$&	\textbf{0.7323}$^\ddag$&	\textbf{0.8967}$^\ddag$	&\textbf{0.8145}$^\ddag$&0.7300$^\ddag$&0.8828$^\ddag$&	0.8067$^\ddag$&	\textbf{0.7268}$^\ddag$	&0.8670$^\ddag$	&\textbf{0.7969}$^\ddag$\\
PLE+PUB & \textbf{0.7142}$^\ddag$&	0.8726$^\ddag$	&\textbf{0.7933}$^\ddag$&	0.7318$^\ddag$&	0.8965$^\ddag$	&0.8142$^\ddag$	&\textbf{0.7307}$^\ddag$&	\textbf{0.8847}$^\ddag$	&\textbf{0.8077}$^\ddag$&	0.7266$^\ddag$&	\textbf{0.8691}$^\ddag$&\textbf{	0.7979}$^\ddag$ \\
\bottomrule
\end{tabular}}
\vspace{-4mm}
\end{table*}

For experiments in NYUV2. We trained each MTO method with MTAN for 300 epochs using the Adam optimizer. Learning rate is 1e-4. batch size is 4, and 10\% steps for warm-up. Different from CAGrad~\cite{liu2021cagrad}, we use task weighted average improvement $\Delta m\%$ as the metric of overall performance.  We first compute average improvement of MTL methods within a task with respect to STL baseline: $\Delta m_{task} = \frac{1}{\#p} \sum_{p} (-1)^{U_{p}} (M_{m,p}-M_{STL,p})/M_{STL,p}$, where $p$ is the criterion of task and $I_{p}=0$ if a lower value better for $p$. Subsequently, the task weighted average improvement is computed as $\Delta m = \frac{1}{3} (\Delta m_{segment}+\Delta m_{depth}+\Delta m_{surface})$. 

\subsection{Two tasks experiments in Recommendation Ranking}

We first report results in the ranking datasets, which are shown in Table~\ref{tab:two_tasks}. We report the result with best average AUC, and all results are averages over 5 times training with different random seeds. AUC improvement on the third decimal place can be considered as significant improvement for offline CTR or CTCVR prediction.
We can draw several insightful conclusions: 
\begin{enumerate}
    \item PUB significantly improves average AUC with both MMOE and PLE backbones in all 4 datasets, while only 2 out of 80 average AUC of baselines outperform MTL backbones. These results demonstrate that PUB largely alleviates seesaw problem~\cite{tang2020ple} in ranking datasets.
    \item Besides, PUB outperforms SOTA baselines at all task metrics. These findings demonstrate the generality and effectiveness of our proposed method. In the Table~\ref{tab:two_tasks}, we observe that some MTL backbones achieve the similar results at certain task metrics, but this actually presents the seesaw problem. Consequently, these backbones exhibit poor overall performance. 
    \item PUB exhibits better robustness against unbalanced loss compared to other baselines. Although, on average, the loss of CTR is approximately 23 times larger than CTCVR, our method still reliably boosts backbones. Conversely, scale-free methods such as NashMTL and IMTL-G, are unable to consistently improve backbones.
\end{enumerate}

\textbf{PUB with UMMs}. In real-world applications, there may be some specific requirements besides alleviating the seesaw problem, such as improving CTCVR more for online advertising platforms. UMMs are commonly used methods to satisfy such special requirements. Conventional MTO methods are unable to integrate UMMs, since they are all based on gradient level tasks fusion or loss level tasks fusion. Fortunately, PUB is flexible and can integrate UMMs easily to accommodate such real-word special requirements. We will assess its capability in the upcoming experiment.

\begin{table}[tp]
\caption{PUB with UMM Experimental Result in Aliexpress\_ES. AUC values are averages over 5 random seeds and only keep 4 decimal. The best results in each column are bold. The $^\ddag$ indicates the result is a significantly improvement than backbone with $p<0.05$. Task weights were searched every 10 steps.}\label{tab:umm_exp}
\centering
\vspace{2mm}
\begin{tabular}{l|cccc}
\toprule
Method&CTR&CTCVR&Average\\\hline
MMOE&0.7290&0.8959&0.8124\\
MMOE+PUB&0.7316$^\ddag$ & 0.8952 & 0.8134$^\ddag$\\
MMOE+PUB+Clippy&\textbf{0.7329}$^\ddag$&0.8954& \textbf{0.8145}$^\ddag$\\
MMOE+PUB+AdaTask&0.7320&\textbf{0.8962}$^\ddag$&0.8141$^\ddag$\\\hline
PLE&0.7306&0.8930&0.8118\\
PLE+PUB&0.7316$^\ddag$ & 0.8946$^\ddag$ & 0.8131$^\ddag$\\
PLE+PUB+Clippy&0.7327$^\ddag$&0.8942$^\ddag$&0.8135$^\ddag$\\
PLE+PUB+AdaTask&0.7315$^\ddag$&0.8959$^\ddag$&0.8137$^\ddag$\\
\bottomrule 
\end{tabular}
\vspace{-4mm}
\end{table}

To demonstrate flexibility of our proposed method, we evaluate PUB with two SOTA update manipulation methods, Clippy~\cite{clippy23} and AdaTask~\cite{yang22adatask}. We re-implement Clippy using Adam according to its paper based on the source code of Fairseq~\cite{ott2019fairseq}. It should be noted that current UMMs are based on certain assumptions, so they may be ineffective in some datasets which do not meet the assumptions. Besides, these methods often require significant effort to adjust hyperparameters and certainly result in larger training cost. Thus, to save costs, we only comprehensively evaluate PUB in the AliExpress\_ES dataset. We search tasks weights every 10 steps in training. Table~\ref{tab:umm_exp} demonstrates the experimental results of PUB with different UMMs. We did not conduct experiments that integrating UMMs with conventional MTO methods, since these methods are hard to combine UMMs because of their fundamental design.

Integrating update manipulation methods, PUB is able to further improve MMOE and PLE in AliExpress\_ES. AdaTask improves PUB at CTCVR task and achieves best result 0.8962 without compromising the performance on the CTR task. Clippy improves PUB at CTR task without hurting CTCVR task too, achieving best result 0.7329. In summary, PUB demonstrates remarkable adaptability when combined with UMMs, and people can select appropriate UMMs based on the specific situation. For fair comparison and saving training cost, we searched task weights every 10 batches on all experiments in above experiments.

\subsection{Experiment on Time Comsuming}\label{sec:time}

To investigate the complexity of PUB, the average one epoch training time is shown on Table~\ref{tab:time}. AliExpress\_ES dataset contains 22326719 training samples. We searched task weights every 10 steps and batch size is 4096. 

\begin{table}[htp]
\caption{The average one epoch training time in AliExpress\_ES. The unit is second and the results are statistic average of all epochs in 5 times training.}
    \label{tab:time}
    \centering
    \begin{tabular}{c|c|c|c}
    \toprule
         Model&Base&NashMTL&PUB  \\\hline
         MMOE&558.26&702.31&651.36\\
         PLE&844.24&1050.65&984.25\\ 
    \bottomrule
    \end{tabular}
    \vspace{-3mm}
\end{table}

As displayed in Table~\ref{tab:time}, the average one epoch training time of PUB is significantly shorter than NashMTL, due to our efficient solution via a sequence of tractable sub-problems that we designed. The speed of PUB is 52.94\% and 67.9\% higher than NashMTL with MMOE and PLE, respectively. All MTO methods need to search task weights during training, so it certainly leads to longer training time. The experiment result shows that PUB is more effective than some SOTA methods. We thought the extra training cost of PUB is acceptable, since the training time is increased only by 16\% and the average weight searching time is 0.17s for MMOE and 0.256s for PLE. Our proposed method can reduce the searching frequency to further save time. According to our experiments, we do see a noticeable drop in overall performance when decrease the searching frequency to every 100 steps.

\subsection{Industrial Evaluation}

We further deployed PUB on a real-world commercial platform \textit{HUAWEI AppGallery}, with hundreds of millions of active users every month, to investigate its effectiveness in a practical setting. The compared baseline is denoted as $\mathcal{M}_{base}$, which is a highly-optimized multi-tasks ranking model over years. $\mathcal{M}_{base}$ adopts Adam optimizer to update trainable parameters. We equip it with PUB algorithm and denote it as $\mathcal{M}_{pub}$. For online A/B testing on a key channel, 5\% of the traffic are randomly selected as the control group and receive results from $\mathcal{M}_{base}$, while another 5\% of the traffic are in the experimental group and receive results from $\mathcal{M}_{pub}$. With continuous improvement, the proportion of $\mathcal{M}_{pub}$ traffic is gradually increased to 20\%, 50\% and eventually \textbf{100\%}. The North Star metric is effective cost per mille (eCPM), which is the result of dividing the ad revenue per banner or campaign by the number of thousand ad impressions. From Table~\ref{tab:industrial}, we observe that with the equipment of PUB algorithm, the $\mathcal{M}_{pub}$ model outperform the baseline method on both eCPM and CTR metircs, which increase revenue for the platform and enhance the experience for the users.

\begin{table}[htp]
\caption{The results of the online A/B testing (eCPM) for our $\mathcal{M}{pub}$ model, compared to the baseline $\mathcal{M}{base}$, were observed over a one-week period with 50\% of the traffic allocated to $\mathcal{M}_{pub}$.}\label{tab:industrial}
    \centering
    \begin{tabular}{c|c|c|c|c}\toprule
          &day1&day2&day3&day4 \\
         \hline
         $\mathcal{M}_{base}$&-&-&-&- \\
         $\mathcal{M}_{pub}$&+3.29\%&+3.35\%&+6.10\%&+4.79\%\\
         \midrule
         &day5&day6&day7&average \\
         $\mathcal{M}_{base}$&-&-&-&- \\
         $\mathcal{M}_{pub}$&+3.70\%&+3.53\%&+3.99\%&+4.06\% \\
         \bottomrule
    \end{tabular} 
    \vspace{-4mm}
\end{table}

\begin{table*}[htp]
\caption{Experimental Result in NYUv2 with Adam optimizer. The results in the table are averaged over 3 training with different random seed, and the best $\Delta m\%$ for each training was saved as the final results. Arrows indicate the values are the higher the better (↑) or the lower the better (↓). The best results in each column are bold. The $^\ddag$ indicates the method outperform baselines.}\label{three_tasks}
\centering
\scalebox{1.2}{\begin{tabular}{cccccccccccc}\toprule
\multirow{3}{*}{\textbf{Method}} & \multicolumn{2}{c}{\textbf{Segment}}         & \multicolumn{2}{c}{\textbf{Depth}}    &     \multicolumn{5}{c}{\textbf{ Surface Normal }} &\multirow{3}{*}{\textbf{$\Delta$ m}\%$\downarrow$}\\
\cmidrule(r){2-3} \cmidrule(r){4-5} \cmidrule(r){6-10}
             &\multirow{2}{*}{mIoU$\uparrow$}&\multirow{2}{*}{Pix Acc$\uparrow$} & \multirow{2}{*}{Abs Err$\downarrow$}  & \multirow{2}{*}{Rel Err$\downarrow$} &  \multicolumn{2}{c}{Angle Distance$\downarrow$}    &  \multicolumn{3}{c}{Within $t^{\circ}\uparrow$}   &         \\
             \cmidrule(r){6-7} \cmidrule(r){8-10}
             &              &         &          &         & Mean           & Median  & 11.25    & 22.5   & 30      &         \\\hline
STL          & 38.31        & 63.76   & 0.6754   & 0.2780  & 25.01          & 19.21   & 30.15    & 57.20  & 69.18   &        \\\hline
\textbf{LBM} &&&&&&&&&&\\
LS           & 39.29        & 65.33   & 0.5493   & 0.2263  & 28.15          & 23.96   & 22.09    & 47.50  & 61.08   & -0.29   \\
SI           & 38.45        & 64.27   & 0.5354   & 0.2201  & 27.60          & 23.37   & 22.53    & 48.57  & 62.32   & -0.54   \\
RLW          & 37.17        & 63.77   & 0.5759   & 0.2410  & 28.27          & 24.18   & 22.26    & 47.05  & 60.62   & 0.72   \\
DWA          & 39.17        & 65.31   & 0.5510   & 0.2285  & 27.61          & 23.18   & 24.17    & 50.18  & 62.39   & -0.65   \\
Uncertainty  & 36.87        & 63.17   & 0.5446   & 0.2260  & 27.04          & 22.61   & 23.54    & 49.05  & 63.65   & -0.30    \\
BanditMTL    & 39.65        & 65.03   & 0.5959   & 0.2436  & 29.66          & 25.83   & 20.00    & 44.00  & 57.41   & 1.17 \\\hline
\textbf{GBM} &&&&&&&&&&\\
MGDA         & 30.47        & 59.90   & 0.6070   & 0.2555  & 24.98          & 19.56   & 29.09    & 56.74  & 69.21   & 0.59  \\
PCGrad       & 38.04        & 64.64   & 0.5550   & 0.2325  & 27.41          & 22.80   & 23.86    & 49.83  & 63.14   & -0.36   \\
CAGrad       & 39.81        & 65.49   & 0.5486   & 0.2250  & 26.31          & 21.58   & 25.61    & 52.36  & 65.58   & -1.44   \\
GradDrop     & 39.39        & 65.12   & 0.5455   & 0.2279  & 27.48          & 22.96   & 23.38    & 49.44  & 62.87   & -0.69 \\
IMTL-G       & 39.25        & 65.41  & 0.5436    & 0.2256  & 26.02          & 21.19   & 26.2     & 53.13  & 66.24   & -1.55  \\
NashMTL      & 40.07        & 65.73   & 0.5348   & 0.2179  & 25.26          & 20.08   & 28.4     & 55.47  & 68.15   & -2.43 \\
FAMO         & 38.94        & 64.91   & 0.5476   & 0.2194  & 25.06          & 19.57   & 29.21    & 56.61  & 68.98   & -2.69 \\ \hline
\textbf{UMM} &&&&&&&&&&\\
AdaTask       & 37.77& 65.52& 0.5179& 0.2131& 25.51& 20.05& 28.44& 55.29& 67.73   & -2.28  \\
ClippyAdam       & 43.87 & 68.75 & 0.5239   & 0.2213  & 26.82  & 22.22   & 24.38     & 50.84  & 64.29   & -2.28  \\
\hline
PUB &44.81$^\ddag$& 69.85$^\ddag$ & \textbf{0.4974} & 0.2087$^\ddag$ &26.49& 21.57 & 25.18    & 52.17  & 65.31   & -3.23$^\ddag$ \\
PUB+ClippyAdam& \textbf{45.59}& \textbf{70.32}& 0.5014$^\ddag$& 0.2118$^\ddag$ & 25.95& 21.03 &26.31& 53.38 &66.48&-3.57$^\ddag$\\
PUB+AdaTask&42.31& 66.69&0.5102$^\ddag$& \textbf{0.2020}& \textbf{24.71}& \textbf{19.03}& \textbf{30.17}& \textbf{57.47}& \textbf{69.63}&\textbf{-3.78}\\
\bottomrule
\end{tabular}}
\vspace{-3mm}
\end{table*}

\subsection{Three tasks experiments in scene understanding}

To investigate the generalization of PUB, we also conducted experiments in scene understanding dataset. We should notice that every influential multi-task optimization approach conducts experiments on scene understanding. The experimental results for the three tasks in the scene understanding experiment are presented in Table~\ref{three_tasks}.We re-implement our method and all baselines. All methods are trained three times with different random seeds by using the same protocol as~\cite{navon2022nashmtl}. 

From Table~\ref{three_tasks}, we can draw the following conclusions: 
\begin{enumerate}
    \item PUB achieves a significant improvement at overall performance metric $\Delta m\%$ compared to SOTA. Although MGDA and FAMO outperform other baselines at surface normal, it still experiences seesaw problem. They focus on the task with small gradient magnitude~\cite{liu2021imtl}, resulting in little improvement at $\Delta m\%$. These results not only demonstrate the superiority of our method, but also its great robustness against large different gradient norm.
    \item Clippy enhances PUB at 7 out of 9 metrics, especially at large gradient magnitude tasks segment, resulting in further improvement at overall performance. PUB with Clippy improves segment task largely and achieves best performance at all metrics of the task.
    \item AdaTask improves PUB at task with small gradient magnitude significantly, outperforming SOTA at 6 out of 9 metrics and obtaining the best overall performance, over 1\% improvement at $\Delta m\%$. More importantly, as far as we all know, PUB with AdaTask is the first method outperforming STL at all metrics in NyuV2 datasets.
\end{enumerate}

Above results demonstrate the great generality of our PUB. It consistently improves overall performance not only in ranking datasets but also CV dataset. Furthermore, the results also demonstrate the effectiveness and flexibility of PUB. Integrating Clippy and AdaTask can further PUB and perform similar to the ranking experiment.

\section{Conclusion and Future Work}
In this paper, we focus on multi-task optimization in ranking models with shared parameters. Our goal is not only to identify the causes behind the challenges posed by conventional multi-task optimization methods in addressing the seesaw problem in multi-task ranking, but also to propose an effective solution that efficiently alleviates this issue.

Firstly, we thoroughly investigate the problem to understand why conventional methods fail to mitigate the seesaw problem. Our paper highlights a fundamental flaw in these methods: they assume that optimal joint gradients consistently result in optimal parameter updates with diverse optimizers. We substantiate this with statistical experiments in CTR and CTCVR ranking datasets, revealing significant discrepancies between actual updates and gradients on shared parameters when using Adam optimizer. Secondly, based on our understanding of the limitations of conventional methods in multi-task ranking, we propose PUB, a Parameter Update Balancing method designed to overcome these limitations. PUB is the first work to optimize multiple tasks through parameter update balancing in field of multi-task optimization. Lastly, our extensive experiments demonstrate the effectiveness of our method on both public ranking datasets and real-world commercial online recommendation system.

\textbf{Future work.} While the initial target of PUB is addressing problems in multi-task ranking, our proposed method shows great potential in other multi-task learning tasks, as it outperforms state-of-the-art general MTO baselines on both ranking and scene understanding datasets. Subsequently, we aim to evaluate the effectiveness of PUB across a wide range of multi-task learning tasks and establish theoretical guarantees for convergence. Lastly, we will conduct more exploration in theory to deeply understand our proposed method, such as converge analysis and updating the moment vectors.

\end{document}